\documentclass[aoas,nameyear,dvips]{arximspdf}
\usepackage{graphicx}

\doi{10.1214/10-AOAS341}
\volume{4}
\issue{4}
\pubyear{2010}
\firstpage{1797}
\lastpage{1823}


\begin{document}
\begin{frontmatter}

\title{Liquid chromatography mass spectrometry-based proteomics:
Biological and technological aspects\thanksref{T1}}
\thankstext{T1}{Portions of this work
were supported by the NIH R25-CA-90301 training grant in biostatistics
and bioinformatics at TAMU, the National Institute of Allergy and
Infectious Disease NIH/DHHS through interagency agreement Y1-AI-4894-01,
National Center for Research Resources (NCRR) grant RR 18522, and were
performed in the Environmental Molecular Science Laboratory, a United
States Department of Energy (DOE) national scientific user facility at
Pacific Northwest National Laboratory (PNNL) in Richland, WA. PNNL is
operated for the DOE Battelle Memorial Institute under contract
DE-AC05-76RLO01830.}
\runtitle{Mass spectrometry-based proteomics}

\begin{aug}
\author[A]{\fnms{Yuliya V.} \snm{Karpievitch}},
\author[A]{\fnms{Ashoka D.} \snm{Polpitiya}},
\author[A]{\fnms{Gordon A.} \snm{Anderson}},
\author[A]{\fnms{Richard D.} \snm{Smith}}
\and
\author[B]{\fnms{Alan R.} \snm{Dabney}\corref{}\ead[label=e1]{adabney@stat.tamu.edu}}

\runauthor{Y. V. Karpievitch et~al.}
\affiliation{Pacific Northwest National Laboratory, Pacific Northwest National Laboratory, Pacific Northwest National Laboratory, Pacific Northwest National Laboratory and Texas A\&M University}
\address[A]{Y. V. Karpievitch\\
A. D. Polpitiya\\
G. A. Anderson\\
R. D. Smith\\
Pacific Northwest National Laboratory\\
Richland, Washington 9935\\
USA}
\address[B]{A. R. Dabney\\
Department of Statistics\\
Texas A\&M University\\
College Station, Texas 77843\\
USA}
\end{aug}

\received{\smonth{5} \syear{2009}}
\revised{\smonth{2} \syear{2010}}

\begin{abstract}
Mass spectrometry-based proteomics has become the
tool of choice for identifying and quantifying the proteome of an
organism. Though recent years have seen a tremendous improvement in
instrument performance and the computational tools used, significant
challenges remain, and there are many opportunities for statisticians to
make important contributions. In the most widely used ``bottom-up''
approach to proteomics, complex mixtures of proteins are first subjected
to enzymatic cleavage, the resulting peptide products are separated
based on chemical or physical properties and analyzed using a mass
spectrometer. The two fundamental challenges in the analysis of
bottom-up MS-based proteomics are as follows: (1) Identifying the
proteins that are present in a sample, and (2) Quantifying the abundance
levels of the identified proteins. Both of these challenges require
knowledge of the biological and technological context that gives rise to
observed data, as well as the application of sound statistical
principles for estimation and inference. We present an overview of
bottom-up proteomics and outline the key statistical issues that arise
in protein identification and quantification.
\end{abstract}

\begin{keyword}
\kwd{LC-MS proteomics}
\kwd{statistics}.
\end{keyword}

\end{frontmatter}
\section{Introduction}

The 1990s marked the emergence of genome sequencing and deoxyribonucleic
acid (DNA) microarray technologies, giving rise to the \mbox{``-omics''} era
of research. Proteomics is the logical continuation of the widely-used
transcriptional profiling methodology [\citet{wilkins1996}].
Proteomics involves the study of multiprotein systems in an organism,
the complete protein complement of its genome, with the aim of
understanding distinct proteins and their roles as a part of a larger
networked system. This is a vital component of modern systems biology
approaches, where the goal is to characterize the system behavior rather
than the behavior of a single component. Measuring messenger ribonucleic
acid (mRNA) levels as in DNA microarrays alone does not necessarily tell
us much about the levels of corresponding proteins in a cell and their
regulatory behavior, since proteins are subjected to many
post-translational modifications and other modifications by
environmental agents. Proteins are responsible for the structure, energy
production, communications, movements and division of all cells, and are
thus extremely important to a comprehensive understanding of systems
biology.

While genome-wide microarrays are ubiquitous, proteins do not share the
same hybridization properties of nucleic acids. In particular,
interrogating many proteins at the same time is difficult due to the
need for having an antibody developed for each protein, as well as the
different binding conditions optimal for the proteins to bind to their
corresponding antibodies. Protein microarrays are thus not widely used
for whole proteome screening. Two-dimensional gel electrophoresis (2-DE)
can be used in differential expression studies by comparing staining
patterns of different gels. Quantitation of proteins using 2-DE has been
limited due to the lack of robust and reproducible methods for
detecting, matching and quantifying spots as well as some physical
properties of the gels [\citet{Ongman2005}]. Although efforts have been
made to provide methods for spot detection and quantification [\citet{Morris2008}], 2-DE is not currently the most widely-used technology
for protein quantitation in complex mixtures. Meanwhile, mass
spectrometry (MS) has proven effective for the characterization of
proteins and for the analysis of complex protein samples [\citet{Vitek2007}]. Several MS methods for interrogating the proteome have
been developed: Surface Enhanced Laser Desorption Ionization (SELDI)
[\citet{tang2004}], Matrix Assisted Laser Desorption Ionization
(MALDI) [\citet{Karas1987}] coupled with time-of-flight (TOF) or
other instruments, and gas chromatography MS (GC-MS) or liquid
chromatography MS (LC-MS). SELDI and MALDI do not incorporate online
separation during MS analysis, thus, separation of complex mixtures
needs to be performed beforehand. MALDI is widely used in tissue imaging
[\citet{Caprioli1997}; \citet{Cornett2007}; \citet{stoeckli2001}]. GS-MS or LC-MS allow for online separation of complex samples
and thus are much more widely used in high-throughput quantitative
proteomics. Here we focus on the most widely-used ``bottom-up'' approach
to MS-based proteomics, LC-MS.

\begin{figure}

\includegraphics{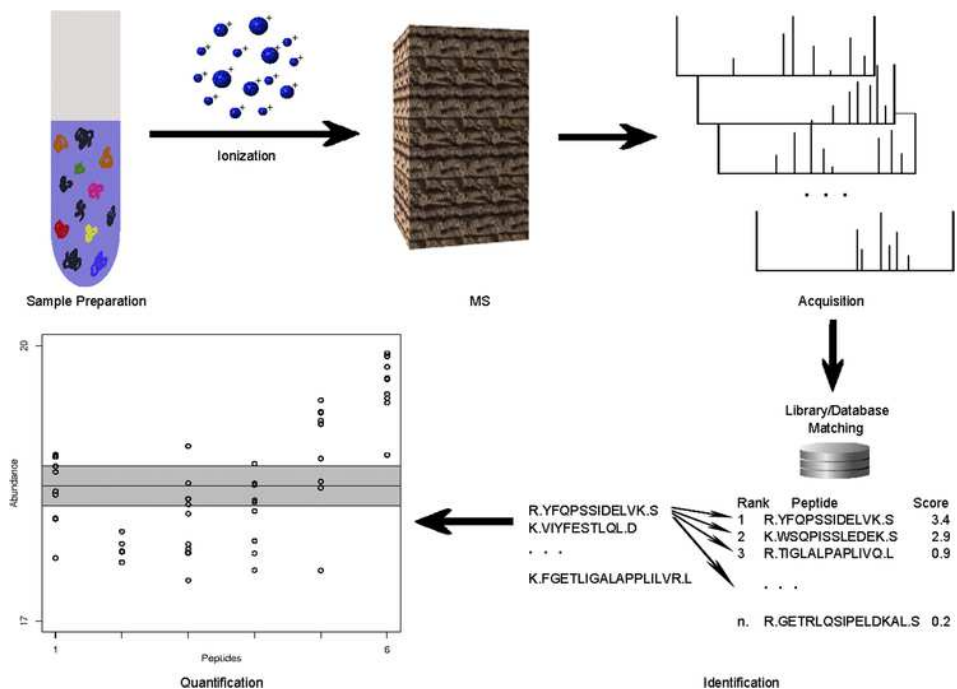}

  \caption{Overview of LC-MS-based proteomics. Proteins are
extracted from biological samples, then digested and ionized prior to
introduction to the mass spectrometer. Each MS scan results in a mass
spectrum, measuring \textit{m/z} values and peak intensities. Based on
observed spectral information, database searching is typically employed
to identify the peptides most likely responsible for high-abundance
peaks. Finally, peptide information is rolled up to the protein level,
and protein abundance is quantified using either peak intensities or
spectral counts.}\label{f1}
\end{figure}

\begin{figure}

\includegraphics{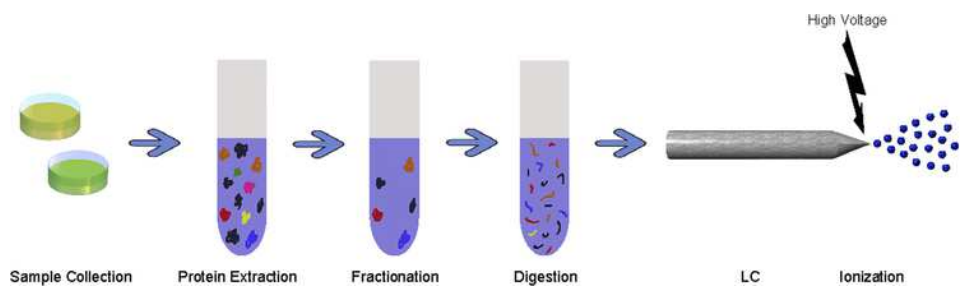}

  \caption{Sample preparation. Complex biological samples are
first processed to extract proteins. Proteins are typically fractionated
to eliminate high-abundance proteins or other proteins that are not of
interest. The remaining proteins are then digested into peptides, which
are commonly introduced to a liquid chromatography column for
separation. Upon eluting from the LC column, peptides are ionized.}\label{f2}
\end{figure}

\begin{figure}

\includegraphics{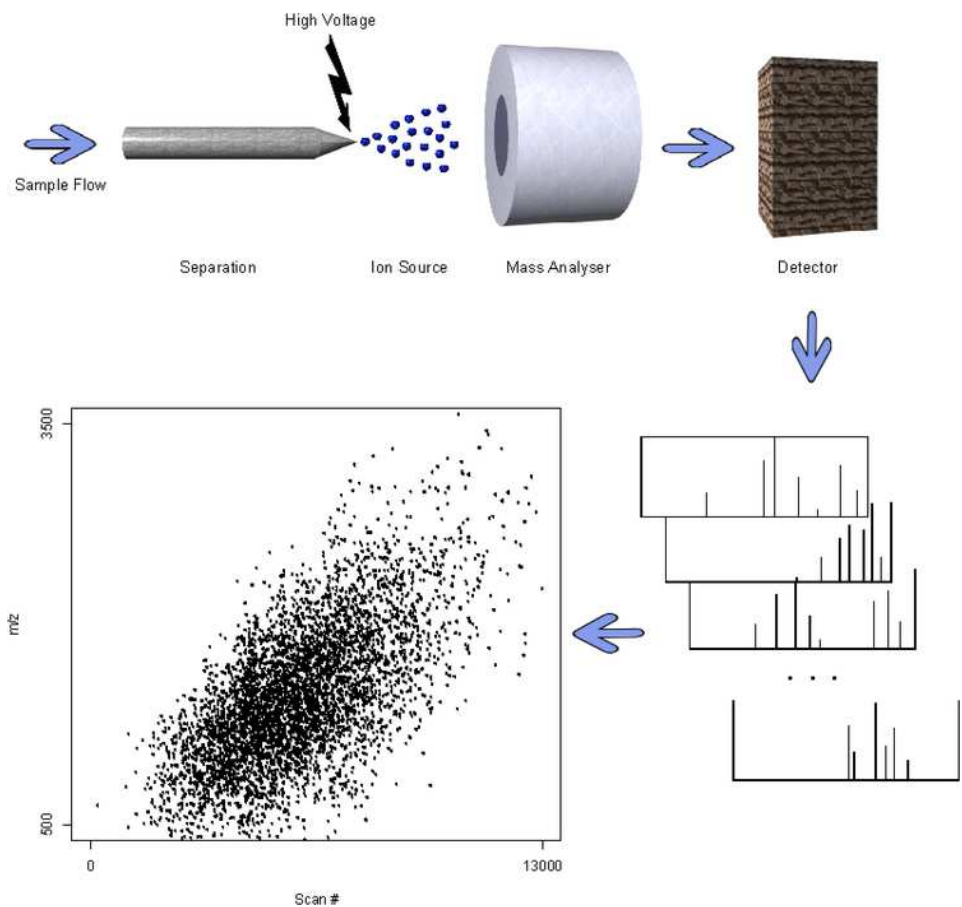}

  \caption{Mass spectrometry. The mass spectrometer consists of
an ion source, responsible for ionizing peptides, the mass analyzer and
the detector, responsible for recording \textit{m/z} values and
intensities, respectively, for each ion species. Each MS scan results in
a mass spectrum, and a single sample may be subjected to thousands of
scans.}\label{f3}
\end{figure}

\begin{figure}

\includegraphics{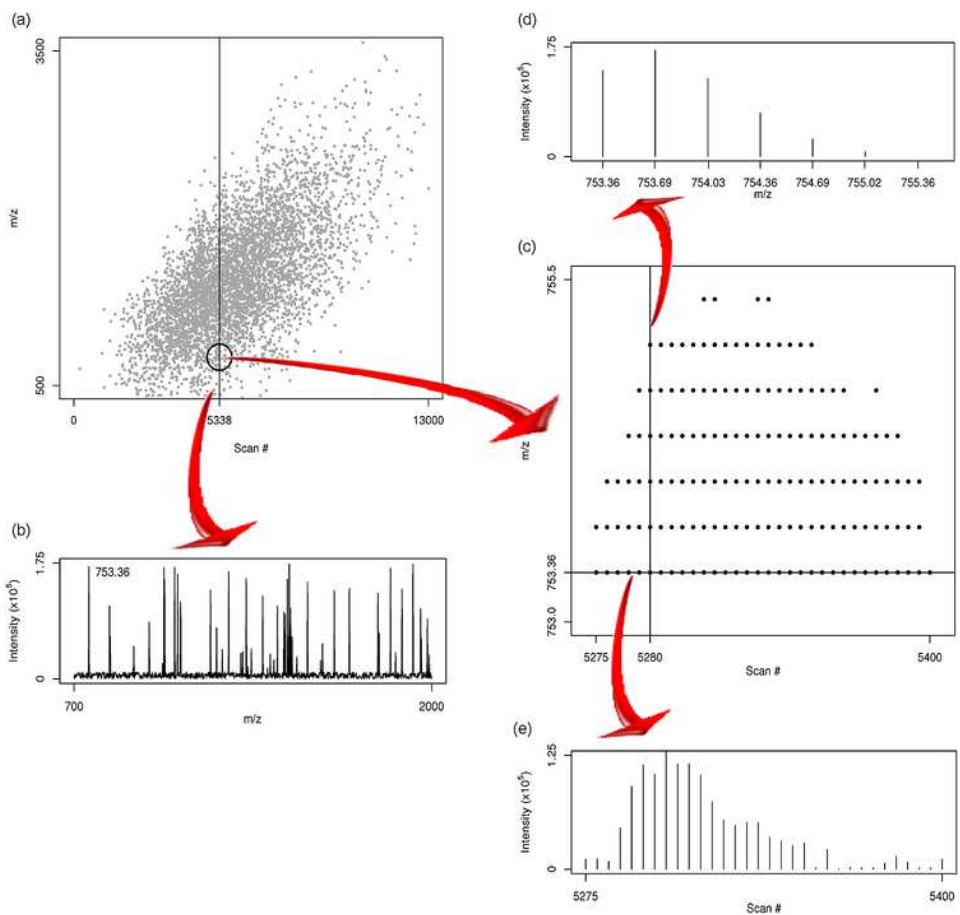}

  \caption{Data acquisition: (\textup{a}) Scan numbers and
\textit{m/z} values for an example raw LC-MS data set. Each individual
scan contains a single mass spectrum. (\textup{b}) The mass spectrum for scan
5338. (\textup{c}) A zoomed-in look at the scans 5275--5400 in \textit{m/z} range
753--755.5. The cluster of dots is indicative of a single LC-MS
``feature.'' (\textup{d}) The isotopic distribution for this feature in scan
5280. Peaks are separated by approximately $1/3$, indicating a charge
state of $+3$. The monoisotopic mass is thus $753.36\times 3 = 2260.08$ Da. (\textup{e})
The elution profile at \textit{m/z} 753.36.}\label{f4}
\end{figure}

\begin{figure}

\includegraphics{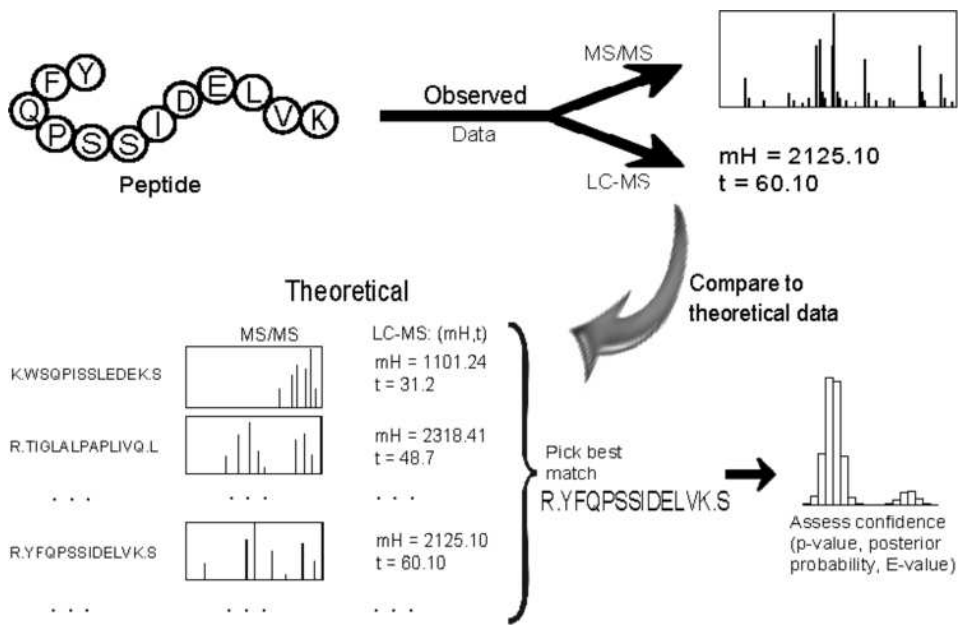}

  \caption{Protein identification. Peptide and protein
identification is most commonly accomplished by matching observed
spectral measurements to theoretical or previously-observed measurements
in a database. In LC-MS/MS, measurements consist of fragmentation
spectra, whereas mass and elution time alone are used in high-resolution
LC-MS. Once a best match is found, one of the following methods for
assessing confidence in the match is employed: decoy databases,
empirical Bayes, or ``expectation values.''}\label{f5}
\end{figure}

\begin{figure}

\includegraphics{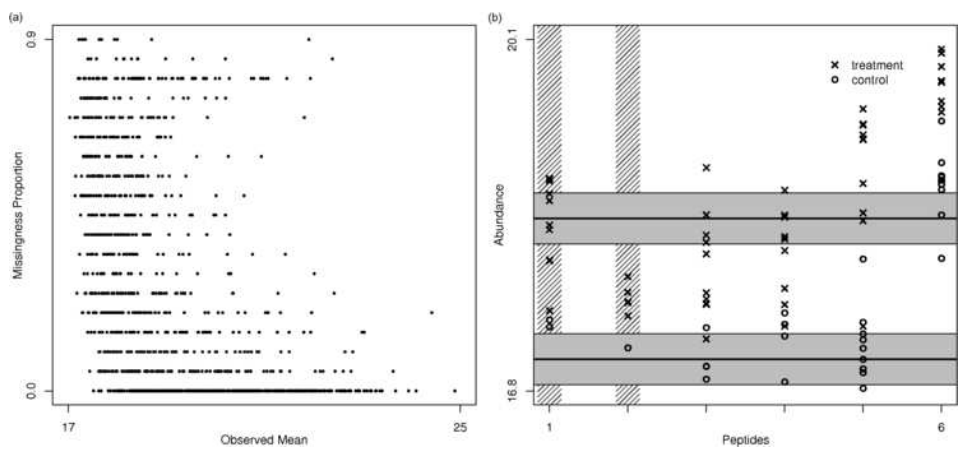}

  \caption{Protein quantitation. The left panel shows the
proportion of missing values in an example data set as a function of the
mean of the observed intensities for each peptide. There is a strong
inverse relationship between these, suggesting that many missing
intensities have been censored. The right panel shows an example protein
found to be differentially expressed in a two-class human study. The
protein had 6 peptides that were identified, although two were filtered
out due to too many missing values (peptides 1 and 2, as indicated by
the vertical shaded lines). Estimated protein abundances and confidence
intervals are constructed from the peptide-level intensities by a
censored likelihood model [Karpievitch et~al. (\protect\citeyear{Karpievitch2009a})].}\label{f6}
\end{figure}

In LC-MS-based proteomics, complex mixtures of proteins are first
subjected to enzymatic cleavage, then the resulting peptide products are
analyzed using a mass spectrometer; this is in contrast to ``top-down''
proteomics, which deals with intact proteins and is limited to simple
protein mixtures [\citet{Han2008}]. A standard bottom-up experiment
has the following key steps (Figures \ref{f1}--\ref{f3}): (a) extraction of proteins
from a sample, (b) fractionation to remove contaminants and proteins
that are not of interest, especially high abundance house-keeping
proteins that are not usually indicative of the disease being studied,
(c)~digestion of proteins into peptides, (d) post-digestion separations
to obtain a more homogeneous mixture of peptides, and (e) analysis by
MS. The two fundamental challenges in the analysis of MS-based
proteomics data are then the identification of the proteins present in a
sample, and the quantification of the abundance levels of those
proteins. There are a host of informatics tasks associated with each of
these challenges (Figures \ref{f4}--\ref{f6}).

The first step in protein identification is the identification of the
constituent peptides. This is carried out by comparing observed features
to entries in a database of theoretical or previously identified
peptides (Figure \ref{f5}). In tandem mass spectrometry (denoted by MS/MS), a
parent ion possibly corresponding to a peptide is selected in MS$^{1}$
for further fragmentation in MS$^{2}$. Resulting fragmentation spectra
are compared to fragmentation spectra in a database, using software like
\mbox{SEQUEST} [\citet{Eng1994}], Mascot [\citet{perkins1999}] or X!Tandem
Alternatively, high-resolution MS instruments can be used to obtain
extremely accurate mass measurements, and these can be compared to mass
measurements in a database of peptides previously identified with high
confidence via MS/MS [\citet{Pasa2004}] using the same software
tools above. In either case, a statistical assessment of the peptide
identification confidence level is desired. Protein identification can
be carried out by rolling up peptide-level identification confidence
levels to the protein level, a process that is associated with a host of
issues and complexities [\citet{Nesvizhskii2003}]. The goal of the
identification process is generally to identify as many proteins as
possible, while controlling the number of false identifications at a
tolerable level. There are a myriad of options for the exact
identification method used, including (i) the choice of a statistic for
scoring the similarity between an observed spectral pattern and a
database entry [\citet{Craig}; \citet{perkins1999}], and
(ii) the choice of how to model the null distribution of the similarity
metric [\citet{Elias2007}; \citet{Keller2002}]. Two other methods
of protein identification exist: de novo and hybrids of de novo and
database matching. This is further explained in
Section~\ref{sec:5}.

In quantitation experiments, protein abundances are inferred from the
identified peptides. One of the most common and simplest methods is to
count the number of times a peptide has been seen and accumulate those
counts for all the peptides seen for a given protein. This gives a value
that is proportional to the abundance of the protein, that is, a more
abundant protein would be expected to have peptides that are observed
more often [\citet{liusadygov2004}; \citet{zhang2009}]. A more accurate
method for quantifying the abundance of a peptide is to calculate the
peak volume (or area) across its elution profile using its extracted ion
chromatogram. Protein abundances are inferred from the corresponding
peptide abundances (Figure \ref{f6}). Peak capacity is a function of the number
of ions detected for a particular peptide, and is related to peptide
abundance [\citet{Old2005}]. Peptide abundances can be computed with
or without the use of stable isotope labels [\citet{Gygi1999}; \citet{wang2003}]. In the case of isotopic labeled experiments, usually a
ratio of the peak capacities of the two isotopically labeled components
is reported. Regardless of the specific technology used to quantify
peptide abundances, statistical models are required to roll
peptide-level abundance estimates up to the protein level. Issues
include widespread missing data due to low-abundance peptides,
misidentified peptides, undersampling of peaks for fragmentation in
MS/MS, and degenerate peptides that map to multiple proteins, among
others. This is further explained in Section \ref{sec:6}.

The purpose of this paper is to provide an accessible overview of
LC-MS-based proteomics. Our template for this paper was a 2002
\textit{Biometrics} paper of similar focus in the DNA microarray setting [\citet{Nguyen2002}]. It is our hope that this, like the 2002 paper for DNA
microarrays, will serve as an entry-point for more statisticians to join
the exciting research that is ongoing in the field of LC-MS-based
proteomics.

\section{Basic biological principles underlying proteomics} \label{sec:2}

Proteins are the major structural and functional units of any cell.
Proteins consist of amino acids arranged in a linear sequence, which is
then folded to make a functional protein. The sequence of amino acids in
proteins is encoded by genes stored in a DNA molecule. The transfer of
information from genetic sequence to protein in eukaryotes proceeds by
transcription and translation. In transcription, single-stranded mRNA
representations of a gene are constructed. The mRNA leaves the nucleus
and is processed into protein by the ribosome in the translation step.
This information transfer, from DNA to mRNA to protein, is essential for
cell viability and function. In genomic studies, microarray experiments
measure gene expression levels by measuring the transcribed mRNA
abundance. Such measurements can show the absence, under- or
over-expression of genes under different conditions. However, protein
levels do not always correspond to the mRNA levels due to a variety of
factors such as alternative splicing or post translational modifications
(PTMs). Thus, proteomics serves an important role in a systems-level
understanding of biological systems.

A three-nucleotide sequence (codon) of mRNA encodes for one amino acid
in a protein. The genetic code is said to be \textit{degenerate}, as
more than one codon can specify the same amino acid. In theory, mRNA
could be read in three different reading frames producing distinct
proteins. In practice, however, most mRNAs are read in one reading frame
due to start and stop codon positions in the sequence. The raw
polypeptide chain (a chain of amino acids constituting a protein) that
emerges from the ribosome is not yet a functional protein, as it will
need to fold into its 3-dimensional structure. In most organisms, proper
protein folding is assisted by proteins called chaperones that stabilize
the unfolded or partially folded proteins, preventing incorrect folding,
as well as chaperonins that directly facilitate folding. Misfolded
proteins are detected and either refolded or degraded. Proteins also
undergo a variety of PTMs, such as phosphorilation, ubiquitination,
methylation, acetylation, glycosylation, etc., which are
additions/removals of specific chemical groups. PTMs can alter the
function and activity level of a protein and play important roles in
cellular regulation and response to disease or cellular damage.

A key challenge of proteomics is the high complexity of the proteome due
to the one-to-many relationship between genes and proteins and the wide
variety of PTMs. Furthermore, MS-based proteomics does not have the
benefit of probe-directed assays like those used in microarrays.
Although protein arrays are available, they (a) are challenging to
design and implement and (b) are not well suited for protein discovery,
and are thus not as widely used as MS-based technologies [\citet{Vitek2007}]. Several steps are involved in preparing samples for MS,
such as protein extraction, fractionation, digestion, separation and
ionization, and each contributes to the overall variation observed in
proteomics data. In addition, technical factors like day-to-day and
run-to-run variation in the complex experimental equipment can create
systematic biases in the data-acquisition stage.

\section{Experimental procedure} \label{sec:3}

A LC-MS-based proteomic experiment requires several steps of sample
preparation (Figure \ref{f2}), including cell lysis to break cells apart,
protein separation to spread out the collection of proteins into more
homogenous groups, and protein digestion to break intact proteins into
more manageable peptide components. Once this is complete, peptides are
further separated, then ionized and introduced into the mass
spectrometer.

\subsection{Sample preparation} \label{sec:3.1}

Analysis of the complete cell proteome usually involves collecting
intact cells, washing them and adding a lysate buffer, containing a
combination of chemicals that break the cell membrane and protease
inhibitors that prevent protein degradation. Cells are homogenized and
incubated with the buffer, after which centrifugation is used to
separate the cellular debris and membrane from the supernatant, or cell
lysate. The cell lysis step is unnecessary when analyzing bodily fluids
such as blood serum. Blood samples are centrifuged, after which red
blood cells pellet at the bottom of the tube, and plasma is collected at
the top. Fibrinogen and other clotting factors are removed to obtain
serum. High abundance proteins are also removed, as usually they do not
play a role in disease. If some of the high abundance proteins are not
removed, they may dominate spectral features and obscure less abundant
proteins of interest. In LC-MS/MS, for example, the most abundant
peptides are selected in the first MS step for further fragmentation in
the following MS step, and only peptides selected for further
fragmentation have a chance to be identified; see Section \ref{sec:3.2} for more
details.

Because of the complexity of the proteome, separation steps are employed
to spread out the proteins according to different chemical or physical
characteristics, making it easier to observe a greater number of
proteins in more detail. At the protein level, two-dimensional gel
electrophoresis (2-DE) is often used to separate on the basis of both
isoelectric point and mass [\citet{Berth2007}; \citet{Gorg2004}].
Proteins in the gel can be stained and extracted. Analyzing each stained
region of the gel separately, for example, would allow for more detailed
assessment of the total collection of proteins in the sample than if all
proteins were analyzed at once. One of the main sources of error in the
gel analysis is unequal precipitation of the proteins between gels.
Thus, horizontal or vertical shifts and even diagonal stretching effects
can be seen in two-dimensional (2-D) gels, necessitating alignment of all
the gels to a reference gel. After gel alignment, spot detection is
performed which may introduce further errors; see Section \ref{sec:7.2} for more
details.

To facilitate protein identification, proteins are usually
cleaved/digested chemically or enzymatically into fragments. Digestion
overcomes many of the challenges associated with the complex structural
characteristics of proteins, as the resulting peptide fragments are more
tractable chemically, and their reduced size, compared to proteins,
makes them more amenable to MS analysis. As examples of digestion
agents, the trypsin enzyme cleaves at the carboxyl side of lysine and
arginine residues, except when either is followed by proline, while
chemical cyanogen bromide (CNBr) cleaves at the carboxyl site of
methionine residues; trypsin is the most commonly used digestive enzyme.
Specificity of the trypsin enzyme allows for the prediction of peptide
fragments expected to be produced by the enzyme and create theoretical
databases. Enzymatic digestion of proteins could be achieved in solution
or gel, although digestion in solution is usually preferred, as gel is
harder to separate from the sample after digestion. Missed cleavages can
cause misidentified or missed peptides when searched against the
database. Database searches can be adjusted to include one or more
missed cleavages, but such searches take longer to complete.

Multiple distinct peptides can have very similar or identical molecular
masses and thus produce a single intense peak in the initial MS
(MS$^{1})$ spectrum, making it difficult to identify the overlapping
peptides. The use of separation techniques not only increases the
overall dynamic range of measurements (i.e., the range of relative
peptide abundances) but also greatly reduces the cases of coincident
peptide masses simultaneously introduced into the mass spectrometer. We
will describe one of the most commonly used separation techniques,
high-performance liquid chromatography (HPLC), which is generally
practiced in a capillary column format for proteomics. Other separation
techniques exist and are similar in that they separate based on some
molecular properties.

A HPLC system consists of a column packed with nonpolar (hydrophobic)
beads, referred to as the stationary phase, a pump that creates pressure
and moves the polar mobile phase through the column and a detector that
captures the retention time. The sample is diluted in the aqueous
solution and added to the mobile phase. As the peptides are pushed
through the column, they bind to the beads proportionally to their
hydrophobic segments. Thus, hydrophilic peptides will elute faster than
hydrophobic peptides. HPLC separation allows for the introduction of
only a small subset of peptides eluting from the LC column at a
particular time into the mass spectrometer. Peptides of similar
molecular mass but different hydrophobicity elute from the LC column and
enter the mass spectrometer at different times, no longer overlapping in
the initial MS analysis. The additional time required for the LC
separation is well worth the effort, as the reduction in the overlap of
the peptides of the same mass in MS$^{1}$ phase dramatically increases
peak resolution (and hence, peak capacity). Note that LC columns must be
regularly replaced, and it is common to observe systematic differences
in the elution times of similar samples on difference columns. Thus,
replacing a column during an experiment may contribute to technical
variation in the resulting observed abundances between two columns.

Further separation techniques include sample fractionation prior to
HPLC, and complementary techniques such as Ion Mobility Separation (IMS)
after HPLC. Multidimensional LC has been successfully used to better
separate peptides. Strong cation exchange (SCX) chromatography is
usually used as a first separation step and reversed-phase
chromatography (RPLC) as a secondary separation step because of its
ability to remove salts and its compatibility with MS through
electrospray ionization (ESI, described below) [\citet{Lee2006};
\citet{Link1999}; \citet{Peng2003}; \citet{Sandra2009}; \citet{sandra2008}]. Combination of SCX with RPLC forms the basis of the
Multidimensional Protein Identification Technology (MudPIT) approach
[\citet{washburn2001}; \citet{wolters2001}]. While
multidimensional LC is capable of achieving greater separation, it
requires larger sample quantities and more analysis time. In HPLC
coupled with IMS, peptides eluting from the HPLC system are ionized
using ESI, and the ions are injected into a drift tube containing
neutral gas at controlled pressure. An electric field is applied, and
the ions separate by colliding with the gas molecules. Larger ions
experience more collisions with the gas and take longer to travel
through the drift tube than smaller ions. IMS is very fast as compared
with HPLC and, when used in conjunction with HPLC, achieves better
separation than HPLC alone. IMS is not entirely orthogonal to HPLC, but
it has been shown to increase the peak capacity (number of detected
peaks) by an order of magnitude [\citet{Belov2007}]. While not
currently in wide use, IMS technologies are rapidly evolving, and
MS-based proteomics will likely involve multiple dimensions of
separation based on both IMS and HPLC in the near future. New algorithms
will need to be developed and existing ones modified to incorporate the
extra separation dimensions.

\subsection{Mass spectrometry} \label{sec:3.2}

A mass spectrometer measures the mass-to-charge ratio (\textit{m/z}) of
ionized molecules. Recent years have seen a tremendous improvement in MS
technology, and there are about 20 different mass spectrometers
commercially available for proteomics. All mass spectrometers are
designed to carry out the distinct functions of ionization and mass
analysis. The key components of a mass spectrometer are the ion source,
mass analyzer and ion detector (Figure \ref{f3}). The ion source is responsible
for assigning charge to each peptide. Mass analyzers take many different
forms but ultimately measure the mass-to-charge (\textit{m/z}) ratio of
each ion. The detector captures the ions and measures the intensity of
each ion species. In terms of a mass spectrum, the mass analyzer is
responsible for the \textit{m/z} information on the \textit{x}-axis, and the
detector is responsible for the peak intensity information on the
\textit{y}-axis.

Ionization methods include electron impact, chemical ionization, fast
atom bombardment, field desorption, electrospray ionization (ESI) and
laser desorption, and they usually operate by the addition of protons to
the peptides. ESI and matrix assisted laser desorption/ionization
(MALDI) are the most widely used methods in proteomics. In the ESI
method, the sample is prepared in liquid form at atmospheric pressure
and flows into a very fine needle that is subjected to a high voltage.
Due to the electrostatic repulsion, the solvent drops leaving the needle
tip dissociate to form a fine spray of highly charged droplets. As the
solvent evaporates, the droplets disappear, leaving highly charged
molecules. ESI is the most effective interface for LC-MS, as it
naturally accommodates peptides in liquid solution. ESI is a soft
ionization method, in that it achieves ionization without breaking
chemical bonds and further fragmenting the peptides. In MALDI analysis,
the biological molecules are dispersed in a crystalline matrix. A UV
laser pulse is then directed at the matrix, which causes the ionized
molecules to eject so that they can be extracted into a mass
spectrometer.

The mass analyzer is key to the sensitivity, resolution and mass
accuracy of an instrument. Sensitivity describes an instrument's ability
to detect low-abundance peptides, resolution to its ability to
distinguish ions of very similar \textit{m/z} values, and mass accuracy
to its ability to obtain mass measurements that are very close to the
truth. There are several basic mass analyzer types: quadrupole (Q),
ion-trap (IT), time-of-flight (TOF), Fourier transform ion cyclotron
resonance (FTICR), and the orbitrap. Different analyzers are commonly
combined to achieve the best utilization as a single mass spectrometer
(e.g., Q-TOF, triple-Q). We do not go into the details of the different
mass analyzer types; interested readers are pointed elsewhere [\citet{Domon2006}; \citet{siuzdak2003}].

In tandem MS (referred to as MS/MS or MS$^{\mathit{n}}$), multiple rounds of MS
are carried out on the same sample. This results in detailed signatures
for detected features, which can be used for identification. Most MS/MS
instruments can automatically select several of the most intense (high
abundance) peaks from a parent MS (MS$^{1})$ scan and subjects the
corresponding ions (precursor or parent ions) for each to further
fragmentation, followed by further scans. This process is repeated until
all candidate peaks of a parent scan are exhausted [\citet{Domon2006}; \citet{zhang2005}]. This results in a fragmentation pattern
for each selected peptide, providing detailed information on the
chemical makeup of the peptide. While the resulting fragmentation
patterns are the basis for identification, MS/MS suffers from
undersampling, in that relatively few (and generally only higher
intensity) precursor ions are selected for fragmentation [\citet{Domon2006}; \citet{Garza}; \citet{zhang2005}]. The
issue of undersampling is not serious enough to steer away from using
MS$^{\mathit{n}}$ for protein identification and quantitation, but researchers
should remember that not all peptides will have equal chances of being
selected for fragmentation and thus may not be observed in the
subsequent MS scans. Furthermore, MS/MS is time-intensive and thus not
always ideal for high-throughput analysis [\citet{Masselon2008}].
Nevertheless, MS/MS is widely used for quantitative MS-based proteomics
and forms the basis for most peptide and protein identification
procedures (Section~\ref{sec:5}). Typically, MS/MS is preceded by LC separation
and can more accurately be denoted by LC-MS/MS.

High-resolution LC-MS instruments (e.g., FTICR) are very fast and can
achieve mass measurements that are sufficiently accurate for
identification purposes. Furthermore, since fragmentation and repeated
scans are not required, the undersampling issues due to peptide
selection for MS/MS are avoided. Still, fragmentation patterns are
valuable for identification, and so hybrid platforms involving both
LC-MS/MS and high-resolution LC-MS are increasingly being used. One such
example is the Accurate Mass and Time (AMT) tag approach [\citet{Pasa2004}; \citet{Tolmachev2008}; \citet{yanofsky2008}]. In the
AMT tag approach, MS/MS analysis is used to create an AMT database of
peptide theoretical mass and predicted elution time, based on
high-confidence identifications from fragmentation patterns, followed by
a single MS run on FTICR to obtain highly accurate mass measurements, as
well as liquid chromatography elution times; peptide identification is
then made by comparing the observed mass measurements and elution times
to the AMT database entries. We note that an AMT database is typically
constructed using many LC-MS/MS runs, resulting in a nearly complete
database of proteotypic peptides [\citet{Mallick2007}]. Because in
the AMT-based approach LC-MS spectra are matched to the database built
from previous multiple MS/MS scans, the undersampling associated with
LC-MS/MS on individual samples is avoided.

\section{Data acquisition}\label{sec:4}

In LC-MS, each sample may give rise to thousands of scans, each
containing a mass spectrum [Figure \ref{f4}(a)]. The mass spectrum for a single
MS scan can be summarized by a plot of \textit{m/z} values versus peak
intensities [Figure \ref{f4}(b)]. Buried in these data are signals that are
specific to individual peptides. As a first step toward identifying and
quantifying those peptides, features need to be identified in the data
and, for example, distinguished from background noise. The first step in
this is MS \textit{peak detection}. Many approaches to peak detection
have been proposed, as this is an old problem in the field of signal
processing. Our lab employs a simple filter on the signal-to-noise ratio
of a peak relative to its local background [\citet{Jaitly2009}]. Each
peptide gives an envelope of peaks due to a peptide's constituent amino
acids. The presence of a peptide can be characterized by the
\textit{m/z} value corresponding to the peak arising from the most
common isotope, referred to as the \textit{monoisotopic} mass. While
there are several isotopes of the elements that make up amino acids,
$^{13}$C is the most abundant, constituting about 1.11\% of all carbon
species. Since the mass difference between $^{13}$C and $^{12}$C is
approximately 1 Da, the monoisotopic peak for a peptide will be
separated from an isotope with a single $^{13}$C by approximately $1/z$,
where $z$ is the charge state of that peptide. Similarly, isotopes with
additional copies of $^{13}$C will be separated in units of
approximately $1/z$. [Figure \ref{f4}(d)].

The process of \textit{deisotoping} a spectrum is often used to simplify
the data by removing the redundant information from isotopic peaks and
involves (i) locating isotopic distributions in a MS scan, (ii)
computing the charge state of each peptide based on the distance between
the peaks in its isotopic distribution, and (iii)~extracting each
peptide's monoisotopic mass. Note that this step is only possible if
sufficiently high-resolution mass measurements have been obtained, as
otherwise isotopic peaks can not be resolved. For (i), detected peaks
are considered as possible members of an isotopic distribution, and
theoretical isotopic distributions, derived from a database of peptide
sequences, are overlaid with the observed spectra. A measure of fit is
computed, and the peaks are called an isotopic distribution if the fit
is good enough. One of the challenges encountered in deisotoping is the
presence of overlapping isotopic distributions from different peptides.
There are many algorithms available for peak detection and deisotoping,
including commercial software from vendors such as Agilent, Rosetta
Biosoftware and Thermo Fisher. Our lab uses Decon2LS [\citet{Jaitly2009}], open-source software that implements a variation of the THRASH
algorithm [\citet{Horn2000}]; the Decon2LS publication contains an
extensive discussion of the above issues, as well as many helpful
references for the interested reader.

A peptide will likely elute from the HPLC over multiple scans, creating
an \textit{elution profile} [Figure \ref{f4}(e)]. Elution profiles for peptides
are typically relatively short in duration, and serve to define a
\textit{feature} in LC-MS data sets. However, there are often
contaminants present in an LC-MS sample with very long elution profiles,
and these are filtered out in preprocessing steps. Various approaches to
summarizing an elution profile are available. Our lab computes a
normalized elution time (NET) [\citet{petritis2006}]. At this stage,
an LC-MS sample has been resolved into a list of LC-MS features, each
with an assigned monoisotopic mass and an elution time. However, due to
mass measurement errors and the random nature of elution times, (mass,
elution time) assigned pairs will vary between LC-MS samples.
\textit{Alignment} is often performed to line up the LC-MS features in
different samples. There are several algorithms for LC-MS alignment;
examples include Crawdad [\citet{Finney2008}] and LCMSWarp [\citet{Jaitly2006}].

As with all high-throughput -omics technologies, MS-based proteomic
data is typically subjected to substantial preprocessing and
normalization. Systematic biases are often seen in mass measurements,
elution times and peak intensities [\citet{Callister2006}; \citet{petyuk2008}]. Filtering of poor-quality proteins and peptides is also
common [\citet{Karpievitch2009a}]. In normalization, care must be
taken to separate biological signal from technical bias [\citet{Dobney2006}]. Widely-used normalization techniques in high-throughput
genomic or proteomic studies involve some variation of global scaling,
scatterplot smoothing or ANOVA [\citet{Quackenbush2002}]. Global scaling
generally involves shifting all the measurements for a single sample by
a constant amount, so that the means, medians or total ion currents
(TICs) of all samples are equivalent. Since common technical biases are
more complex than simple shifts between samples, global scaling is
unable to capture complex bias features. Scatterplot smoothing, TIC and
ANOVA normalization methods are sample-specific and hence more flexible.
However, more complex preprocessing steps can result in overfitting,
causing errors in downstream inference. For example, fitting a complex
preprocessing model may use up substantial degrees of freedom, and
analyzing the processed data, assuming that no degrees of freedom have
been used, may result in overly optimistic accuracy levels and
overestimated statistical significance; specific examples can be seen in
\citet{Karpievitchyv2009b}. Ideally, preprocessing would be carried out
simultaneously with inference, or the downstream inferential steps would
incorporate knowledge of what preprocessing was done [\citet{Leek2007}]. A~recently proposed method, called EigenMS, removes bias of
arbitrary complexity by the use of the singular value decomposition to
capture and remove biases from LC-MS peak intensity measurements
[\citet{Karpievitchyv2009b}]. EigenMS removes biases of arbitrary
complexity and adjusts the normalized intensities to correct the
\textit{p}-values after normalization (ensuring that null \textit{p}-values are uniformly
distributed).

Mass spectrometer manufacturers have developed a variety of proprietary
binary data formats to store instrument output. Examples include
.\textit{baf} (Bruker), \textit{.Raw} (Thermo) and \textit{.PKM}
(Applied Biosystems). Handling data in different proprietary formats
typically requires corresponding proprietary software, making it
difficult to share datasets. Several open-source, XML-based
vendor-independent data formats have recently been developed to address
this limitation: mzXML [\citet{Lin2005}; \citet{Pedrioli2004}],
mzData [\citet{Orchard2009}] and mzML [\citet{Deutsch2008}; \citet{Orchard2007}]. mzML 1.0 was released in June 2009 and is considered a
merge of the best of mzData and mzXML. The format can store spectral
information, instrument information, instrument settings and data
processing details. mzML also has extensions such as chromatograms and
multiple reaction monitoring (MRM) profile capture, and it now replaces
both mzData and mzXML.

\section{Protein identification}\label{sec:5}

In bottom-up proteomics protein identification is usually accomplished
by first comparing observed MS features to a database of predicted or
previously identified features (e.g., by MS/MS or on the basis of
previous analysis of a well characterized sample, Figure \ref{f5}). The most
widely-used approach is tandem MS with database searching [\citet{Vitek2007}], in which peptide fragmentation patterns are compared to
theoretical patterns in a database using software like Sequest [\citet{Eng1994}], X!Tandem [\citet{Craig}] and Mascot [\citet{perkins1999}].
With high-resolution LC-MS instruments, identifications can
be made on the basis of mass and elution time alone, or in conjunction
with MS/MS fragmentation patterns [\citet{Pasa2004}].
Alternatives to database-searching include (i) \textit{de novo} peptide
sequencing [\citet{Dancik1999}; \citet{Johnson2005}; \citet{Lu2003}; \citet{standing2003}] and (ii) hybrids of the \textit{de novo} and
database searching approaches [\citet{Frank2005}; \citet{sunyaev2003}; \citet{tabb2003}; \citet{tanner2005}]. For detailed
reviews of the database searching algorithms see \citet{Kapp2007},
\citet{Nesvizhskii2007}, \citet{Vitek2007}, \citet{sadygov2004} and \citet{yates1998}.

In tandem MS, precursor ions for the most abundant peaks in a scan are
fragmented and scanned again. In collision-induced dissociation (CID),
precursor ions are fragmented by collision with a neutral gas [\citet{Laskin2003}; \citet{pttenauer2009}; \citet{sleno2004}; \citet{wells2005}]. Subsequent MS analysis measures the
\textit{m/z} and intensity of the fragment ions (product or daughter
ions), creating a fragmentation pattern (Figure \ref{f5}). CID usually leads to
\textit{b}- and \textit{y}-ions through breakage of the amide bond along the peptide
backbone. \textit{b}-ions are formed when the charge is retained by the
amino-terminal fragment, and \textit{y}-ions are formed when charge is retained
by the carboxy-terminal fragment. Breaks near the amino acids glutamic
acid (E), aspartic acid (D) and proline (P) are more common, as well as
breaking of the side-chains [\citet{sobott2009}]. Other fragmentation
patterns are possible, such as \textit{a}-, \textit{c}-, \textit{x}- and \textit{z}-types. Electron capture
dissociation (ECD) produces \textit{c}- and \textit{z}-ions and leaves side-chains intact.
The fragmentation pattern is like a fingerprint for a peptide. It is a
function of amino acid sequence and can therefore be predicted. The
observed fragmentation pattern should match well with its theoretical
pattern, assuming that its peptide sequence is included in the search
database.

A search database is created by specifying a list of proteins expected
to contain any proteins present in a sample. In human studies, for
example, the complete known proteome can be specified with a FASTA file,
which can then be used to create peptide fragment sequences by
simulating digestion with trypsin. For each resulting peptide, a
theoretical fragmentation pattern is then created. For details on
protein digestion and fragmentation see \citet{siuzdak2003}. Several software
programs are available for database matching (e.g., SEQUEST, X!Tandem
and Mascot). Each has its own algorithm for assessing the fit between
observed and theoretical spectra, and there can be surprisingly little
overlap in their results [\citet{searle2008}]. Note that a correct
match can only be made if the correct sequence is in the database in the
first place. If an organism's genome is incomplete or has errors, this
will not be the case. Furthermore, because of undersampling issues in
MS/MS, only a small percentage of peptides present in a sample will even
be considered for identification. This is due to the fact that only a
small portion of higher abundance peaks (for example, the 10 most
abundant peaks) are selected from the spectra in the first MS step for
fragmentation in the second MS analysis. Thus, lower abundance proteins
are obscured by the presence of the high abundance ones.

High-resolution LC-MS instruments can be used to identify peptides on
the basis of extremely accurate mass measurements and LC elution times.
A database is again required, containing theoretical or
previously-observed mass and elution time measurements. In hybrid
approaches, like the AMT tag approach [\citet{Pasa2004}],
identifications from MS/MS are used to create a database of putative
mass and time tags for comparison with high-resolution LC-MS data. Since
MS/MS is sample- and time-intensive, hybrid approaches allow for
higher-throughput analysis, subjecting only a subset of the sample to
MS/MS and the rest to rapid LC-MS. Alternatively, previously-observed
MS/MS fragmentation patterns can be used to create a mass and time tag
database. By using many LC-MS/MS datasets in the creation of the
database, the undersampling issues associated with LC-MS/MS are avoided.

In each of the above approaches, there is a statistical problem of
assessing confidence in database matches. This is typically dealt with
in one of two ways. The first involves modeling a collection of database
match scores as a mixture of a correct-match distribution and an
incorrect-match distribution. The confidence of each match is assessed
by its estimated posterior probability of having come from the
correct-match distribution, conditional on its observed score [\citet{Kall2008a}]; PeptideProphet is a widely-used example [\citet{Keller2002}].
Improvements have been made to PeptideProphet to avoid fixed
coefficients in computation of discriminant search score and utilization
of only one top scoring peptide assignment per spectrum [\citet{Ding2008}]. Decoy databases are an alternative approach, in which the
search database is scrambled so that any matches to the decoy database
can be assumed to be false [\citet{Choighost2008}; \citet{Kallet2008b}]. The distribution of decoy matches is then used as the null
distribution for the observed scores for matches to the search database,
and \textit{p}-values are computed as simple proportions of decoy matches as
strong or stronger than the observed matches from the search database. A
hybrid approach that combines mixture models with decoy database search
can also be used [\citet{Choinesvizhskii2008a}]. Whether working from
posterior probabilities or \textit{p}-values, lists of high-confidence peptide
identifications can be selected in terms of false discovery rates [\citet{Nesvizhskiichoi2008b}; \citet{storley2003}]. Both decoy
database matching and empirical Bayes approaches are global, in that
they model the distribution of database match scores for all spectra at
the same time. An ``expectation value'' is an alternative significance
value, which models the distribution of scores for a single experimental
spectrum with all peptide match scores from the theoretical database
[\citet{Fenyo2003}].

An alternative to database search approaches is \textit{de novo}
sequencing [\citet{Dancik1999}; \citet{Frank2005}; \citet{Johnson2005}; \citet{Lu2003}; \citet{standing2003}; \citet{tabb2003}].
\textit{De novo} sequencing involves assembling the amino acid sequences
of peptides based on direct inspection of spectral patterns. For a given
amino acid sequence, the possible fragmentation ions and masses can be
enumerated, as well as the expected frequency with which each type of
fragment ion would be formed. \textit{De novo} sequencing therefore
tries to find the sequence for which an observed spectral pattern is
most likely. The key distinction from database-search approaches is that
there is no need for a priori sequence knowledge. Suppose, for
example, that we are studying human samples. With database-search, we
would load a human proteome FASTA file and only have access to amino
acid sequences generated therein. With \textit{de novo} sequencing,
\textit{any} amino acid sequence could be considered. This can be
important when studying organisms with incomplete or imperfect genome
information [\citet{ram2005}]. Drawbacks include increased
computational expense as well as the need for relatively large sample
quantities.

Combinations of \textit{de novo} sequence tag generation and database
searching (hybrid methods) are widely used in PTM identification [\citet{Mann}]. The \textit{de novo} approach infers a peptide
sequence tag (not the full-length peptide) from the spectrum without
searching the protein database. These sequence tags can then be used to
filter the database to reduce its size, which in turn speeds up the
calculation of the spectrum matches with all possible PTMs. InsPect is a
widely used tool for identification of PTMs [\citet{tanner2005}]. Lui
et~al. proposed a similar sequence tag-based approach with a
deterministic finite automaton model for searching a peptide sequence
database [\citet{Liu2006}].

While bottom-up MS-based proteomics deals with peptides, the real goal
is to identify \textit{proteins} present in a sample. In most cases, a
peptide amino acid sequence can be used to identify the protein from
which it was derived. Software like ProteinProphet can translate
peptide-level identifications to the protein level and assign each
resulting protein identification a confidence measure [\citet{Nesvizhskii2003}]. A key challenge in translating peptide identifications to
the protein level is \textit{degeneracy}. A degenerate peptide is one
that could have come from multiple proteins; this is most common for
peptides with short amino acid sequences or ones that come from
homologous proteins (where homology refers to a similarity in amino acid
sequences). Based on the information present in an individual degenerate
peptide, it is not necessarily clear how to decide between multiple
proteins. However, by taking the information present in uniquely
identified and degenerate peptides that were identified as belonging to
multiple proteins into account, sensible model-based decisions can be
achieved [\citet{shen2008}]. PeptideProphet shares degenerate
peptides among their corresponding proteins and produces a minimal
protein list that accounts for such peptides. Another challenge is due
to the fact that correctly identified peptides usually belong to a small
set of proteins, but incorrectly identified peptides match randomly to a
large variety of proteins. Thus, a small number of incorrectly
identified peptides (with high scores) can make it difficult to
determine the correct parent protein, especially in a single-peptide
identification, and may result in a much higher error rate at the
protein level [\citet{Nesvizhskii2004}].

\section{Protein quantitation}\label{sec:6}

Quantitative proteomics is concerned with quantifying and comparing
protein abundances in different conditions (Figure \ref{f6}). There are two
main approaches: stable isotope labeling and label free. In all cases,
as in the identification setting, there is the challenge of rolling
peptide-level information up to the protein level. This can be viewed as
an analogous problem to the probe-set summarization step required with
many DNA microarrays [\citet{Li2001}].

In label-based quantitative LC-MS, chemical, metabolic or enzymatic
stable isotope labels are incorporated into control and experimental
samples, the samples are mixed together and then analyzed with LC-MS
[\citet{Goshe2003}; \citet{Guerrera}; \citet{Gygi1999}]. In chemical labeling, such as isotope-coded affinity tag
(ICAT), Cystine (\textit{Cys}) residues are labeled [\citet{Gygi1999}]. In metabolic labeling, cells from two different conditions are
grown in media with either normal amino acids
($^{1}$H/$^{12}$C/$^{14}$N) or stable isotope amino acids
($^{2}$H/$^{13}$C/$^{15}$N) [\citet{Oda1999}; \citet{Ong2002}].
This approach is not applicable to human or most mammalian protein
profiling. In enzymatic labeling, proteins from two groups are digested
in the presence of normal water (H$_{2}{}^{16}$O) or isotopically labeled
water (H$_{2}{}^{18}$O) [\citet{schnolzer1996}; \citet{ye2009}]. In
all of the above methods, differences in label weight create a shift in
\textit{m/z} values for the same peptide under the two conditions. After
tandem mass analysis (LC-MS/MS), spectra are matched against a database,
and ratios of peptide abundances in the two conditions are determined by
integrating the areas under the peaks of each labeled ion that was
detected. Strong linear agreement has been shown between true
concentrations and those estimated by label-based approaches [\citet{Old2005}]. Of the two quantitation methods considered here,
label-based methods are able to achieve the most precise estimates of
relative abundance. Limitations include the following: (i) its
restriction to two comparison groups, (ii) associated difficulties with
incorporating future samples into an existing data set, and (iii)
expense. A newer method that allows for the comparison of four treatment
samples at a time and avoids the cystine-selective affinity of ICAT is
iTRAQ [\citet{ross2004}; \citet{thompson2003}; \citet{wiese2007}]. iTRAQ uses isobaric labels at N-terminus which have two
components: reporter and balance moieties. Combined reporter and balance
moieties always have masses of 145 Da. For example, if for treatment
group one we use reporter of mass 114 and balance of mass 31, then for
another treatment group we can use reporter of mass 116 and balance of
mass 29. Precursor ions from all treatment groups appear as a single
peak of the same weight in MS$^{1}$. After further fragmentation,
peptides break down into smaller pieces and separate balance and
reporter ions. Reporter ions thus appear as distinct masses, and peptide
abundances are determined from those. iTRAQ is limited to four or eight group
comparisons, but limitations (ii) and (iii) above still apply.

Label-free quantitative analysis measures relative protein abundances
without the use of stable isotopic labels. In contrast to label-based
methods, samples from different comparison groups are analyzed
separately, allowing for more complex experiments as well as the
addition of subsequent samples to an analysis; label-free methods are
also faster than label-based methods. Label-free quantification can be
grouped into two categories: spectral feature analysis and spectral
counting. In spectral feature analysis, peak areas of identified
peptides are used for abundance estimates. The peak areas are sometimes
normalized to the peak area of an internal standard protein spiked into
the sample at a known concentration level. Good linear correlation
between estimated and true relative abundances has been shown for this
method of peptide quantification [\citet{Bondarenko2002}; \citet{Chelius2002}; \citet{Old2005}; \citet{wang2003}]. In spectral
counting, peptide abundances for one sample are estimated by the count
of MS/MS fragmentation spectra that were observed for each identified
peptide [\citet{Choi2008}]. Repeated identifications of the same
peptide in the same sample are due to its presence in several proximal
scans constituting its elution profile. Good linear correlation between
true and estimated relative abundance from spectral counting have been
shown [\citet{Ghaemmaghami2003}; \citet{liusadygov2004}]. Spectral counts
are easy to collect and do not require peak area integration like
spectral peak analysis or label-based methods.

Missing peptides are common in MS-based proteomic data. In fact, it is
common to have 20--40\% of all attempted intensity measures missing.
Abundance measurements are missed if, for example, a peptide was
identified in some samples but not in others. This can happen in several
ways: (i) the peptide is present in low abundances, and in some samples
the peak intensities are not high enough to be detected or for the
corresponding ions to be selected for MS/MS fragmentation, (ii)
competition for charge in the ionization process, by which some ion
species are liable to be dominated by others, and (iii) peptides whose
chemical or physical structure cause them to get trapped in the LC
column, among others. Mechanism (i) is essentially a censoring mechanism
and appears to be responsible for the vast majority of missing values
[Figure \ref{f6}(a)]. This complicates intensity-based quantitation, as simple
solutions will tend to be biased. For example, analysis of only the
observed intensities will tend to overestimate abundances and
underestimate variances. Simple imputation routines like row-means or
\textit{k}-nearest-neighbors suffer from similar limitations. Statistical models
are needed to address these issues, as well as to handle the
peptide-to-protein rollup [\citet{Karpievitch2009a}; \citet{wang2006}; also, see Figure \ref{f6}(b)]. Note that a further benefit of spectral
counting is that it is less sensitive to missing values.

We note that protein identification and quantitation are complementary
exercises. Unidentified proteins cannot be quantified, and the
confidence with which a protein was identified should perhaps be
incorporated into that protein's abundance estimate. Degenerate
peptides, for example, present problems for both identification and
quantitation, but evidence for the presence of sibling peptides from one
protein in high abundance can be useful in deciding between multiple
possible protein identities.

\section{Other technologies}\label{sec:7}

\subsection{MALDI and mass fingerprinting} \label{sec:7.1}

MALDI (matrix assisted lazed desorption ionization) is mostly used for
single MS, typically using a TOF mass analyzer. MALDI refers to the
method of ionization, in which a laser is pulsed at a crystalline matrix
containing the sample (analyte) [\citet{Guerrera}; \citet{Karas1987}]. The analyte is mixed with the matrix solution, spotted in a
well on a MALDI plate and allowed to crystallize. The matrix consists of
small organic molecules that absorb light at the wavelength of the laser
radiation. Upon absorption, the matrix molecules transfer energy to the
sample molecules to permit ionization and desorption of even large
molecules as intact gas-phase ions; the matrix also serves to protect
the analyte from being destroyed by the laser pulse. MALDI is considered
a soft ionization technique, resulting in very little analyte
fragmentation. Crystallized samples can be stored for some time before
analysis or for repeated analysis.

While MALDI MS/MS instruments exist, MALDI is most commonly used for
mass fingerprinting, where spectral patterns are identified for
discriminating samples from different conditions (e.g., cancer vs.
normal). Machine learning techniques, such as linear discriminant
analysis, Random Forest and Support Vector Machine, among others, are
typically used to build classifiers in hopes of finding tools for the
early detection of a disease. Disease biomarkers (specific \textit{m/z}
values) can be identified from the set of the differentially expressed
features. However, to date, the success rate for identification of true
biomarkers is low, in part due to the poor reproducibility of the
experiments in time and between labs [\citet{Baggerly2004}; \citet{petricoin2002}].

\subsection{2-D gels} \label{sec:7.2}

2-D gel electrophoresis (2-DE) is an alternative technique for protein
separation [\citet{Gorg2004}; \citet{Kobalz1995}; \citet{weiss2009}], first introduced in 1975 [\citet{Klose1975}; \citet{OFarrell1975}].
Here, two orthogonal separations are used: proteins are first separated
based on their isoelectric point (pI), then based on their size (mass).
The first dimension utilizes the fact that the net charge of the protein
is pH-dependent. Proteins are loaded into the pH gradient (variable pH)
and subjected to high voltage. Each protein migrates to the pH location
in the gradient where its charge is zero and becomes immobilized there.
The second dimension gel contains SDS, detergent molecules with
hydrophobic tails and negatively charged heads. SDS denatures (unfolds)
the proteins and adds negative charge in proportion to the size of the
protein. An electric field is applied to move negatively charged
proteins toward the positively charged electrode, smaller proteins
migrating through the gel faster than larger ones. Multiple copies of
the proteins will generally move at the same speed and will end up
fixated in bulk at a certain spot on the gel.

Protein detection is performed with staining (most common) or
radio-labeling. Proteins can then be quantified based on their spot
intensity. The staining intensity is approximately a linear function of
the amount of protein present. Images of the 2-D gels can be compared
between different comparison groups to study protein variations between
the groups and identify biomarkers. The following steps are generally
required before quantitative and comparative analysis can be done, not
necessarily in this order: (a) denoising, (b) background correction, (c)
spot detection, (d) spot matching/gel alignment, (e) spot
quantification. Although all steps are needed, spot matching is the most
important, as proteins can shift along the axis from image to image (gel
to gel) as well as exhibit a pattern of stretching along the diagonals.
Examples of programs designed to perform the above steps are Progenesis
(Nonlinear Dynamics Ltd., Newcastle-upon-Tyne, UK) and PDQuest Version
8.0 (Bio-Rad Laboratories, Hercules, CA, USA), both of which are
proprietary. Pinnacle is an open source program that performs spot
detection and quantification in the aligned gels [\citet{Morris2008}].

\section{Discussion}\label{sec:8}

While the field of LC-MS-based proteomics has seen rapid advancements in
recent years, there are still significant challenges in proteomic
analysis. The complexity of the proteome and the myriad of computational
tasks that must be carried out to translate samples into data can lead
to poor reproducibility. Advancements in mass spectrometry and
separation technologies will surely help, but there will continue to be
a crucial role for statisticians in the design of experiments and
methods specific to this setting. Careful assessments of the
capabilities of current LC-MS-based proteomics to achieve certain levels
of sensitivity and specificity, based on instrument configuration,
experimental protocol, experimental design, sample size, etc., would be
extremely valuable for assisting in the establishment of best-practices,
as well as for gauging the capabilities of the technology; the National
Cancer Institute's Clinical Proteomic Technologies for Cancer program is
an example. It is likely the case that very large studies will be
required for true breakthrough findings (e.g., biomarkers) in systems
biology using proteomics.

Specific methodological areas that can use additional input from
statisticians include the development of statistical models for rolling
up from peptides to proteins; determination of protein networks;
construction of confidence levels with which we identify peptides and
subsequently proteins; alignment of LC-MS runs and assurance of quality
of those alignments, that is, assigning a \textit{p}-value to a set of aligned
LC-MS runs to assess ``correctness'' of alignment. Furthermore, as
additional dimensions of separation (as in IMS-LC-MS) are introduced,
more flexible and generalizable preprocessing, estimation and
inferential methods will be required. In general, statisticians can play
a pivotal role in LC-MS-based proteomics (as well as other -omics
technologies) by participating in interdisciplinary research teams and
assisting with the application of classical statistical concepts [\citet{Oberg2009}]. In particular, the statistician can contribute by
ensuring that well-planned experimental designs are employed,
assumptions required for reliable inference are met, and proper
interpretation of statistical estimates and inferences are used
[\citet{Dougherty2009}; \citet{Hand2006}]. These contributions are arguably more
valuable than the development of additional algorithms and computational
methods. Due to the great complexity of high-throughput -omics
technologies and the data that result, careful statistical reasoning is
imperative.

\section*{Acknowledgment}

We thank Josh Adkins for many helpful discussions.

\printaddresses

\end{document}